\documentstyle[12pt,psfig]{article}

\begin{document}

\begin{flushleft}
{\LARGE \bf Average persistence in random walks} \\[5mm]
\end{flushleft}
\null\hspace*{1cm}
\begin{minipage}[t]{15cm}
{\large Heiko~Rieger$(*)$ and Ferenc Igl\'oi$(**)$ } \\[5mm]
{\it $(*)$ HLRZ, Forschungszentrum J\"ulich, 52425 J\"ulich, Germany\\
Institut f\"ur Theoretische Physik, Universit\"at zu K\"oln, 50923 K\"oln, 
Germany} \\
{\it $(**)$ Research Institute for Solid State Physics and Optics, 
H-1525 Budapest, P.O.Box 49, Hungary\\ Institute for Theoretical Physics,
Szeged University, H-6720 Szeged, Hungary} 
\\[5mm]
\today \\[5mm]
probability, anomalous diffusion in disordered environments,
random quantum spin chains\\[8mm]
{\small
{\bf Abstract. ---}
We study the first passage time properties of an integrated Brownian
curve both in homogeneous and disordered environments. In a disordered
medium we relate the scaling properties of this center of mass
persistence of a random walker to the average persistence, the latter
being the probability $\overline{P}_{\rm pr}(t)$ that the expectation
value $\langle x(t)\rangle$ of the walker's position after time $t$
has not returned to the initial value. The average persistence is then
connected to the statistics of extreme events of homogeneous random
walks which can be computed exactly for moderate system sizes. As a
result we obtain a logarithmic dependence $\overline{P}_{\rm
  pr}(t)\sim \ln(t)^{-\overline{\theta}}$ with a new exponent
$\overline{\theta}=0.191\pm0.002$. We note on a complete
correspondence between the average persistence of random walks and the
magnetization autocorrelation function of the transverse-field Ising
chain, in the homogeneous and disordered case.}\\[8mm]
\end{minipage}

\newcommand{\bc}{\begin{center}}
\newcommand{\ec}{\end{center}}
\newcommand{\be}{\begin{equation}}
\newcommand{\ee}{\end{equation}}
\newcommand{\beqn}{\begin{eqnarray}}
\newcommand{\eeqn}{\end{eqnarray}}

\newcommand{\deffig}[5]{
\begin{figure}[t]
\begin{minipage}[t]{8cm}
\caption[*]{{\small #5}}
\label{#1}
\end{minipage}
\begin{minipage}[t]{8cm}
\null\ 
\protect{\psfig{figure={#2},height={#3},width={#4}}}
\end{minipage}
\end{figure}
}

\parskip=0cm

First passage time or persistence problems have a long history in the
physical literature \cite{vankampen}. Recently they gained a lot of
interest, since the persistence exponents, describing the asymptotic
behavior of first passage time probabilities, are shown to be
independent dynamical critical exponents, which have been calculated
for various models exactly\cite{derrida,persistence}. Not much is
known about analogous quantities in systems with quenched disorder,
for instance the random walk (or diffusion) in a disordered
environment \cite{review}, which is in the one-dimensional case the
Sinai-model\cite{sinai}. For this model the first passage /
persistence exponent for a single walker has been determined by us in
a previous work \cite{diffusion}. In this letter we introduce and
study the concept of {\it average} persistence of random walks both in
homogeneous and random environments.

We consider a random walk with nearest neighbor hopping in one
dimension defined by the Master equation
\be
p_i(t=0)=\delta_{i,1}\;,\quad
\frac{d}{dt} p_i(t) =
-(w_{i, i-1}+w_{i, i+1})p_i(t)
+w_{i-1, i}p_{i-1}(t)
+w_{i+1, i}p_{i+1}(t)
\label{master}
\ee
describing the time evolution of the probability $p_i(t)$ for the
walker to be at site $i$ after time $t$ when having been initially at
site $i=1$. The homogeneous random walk is defined via uniform
transition rates $w_{i,i+1}=w_{i+1,i}=1/2$ and the random walk in a
disordered environment is modeled by choosing the transition rates to
be quenched random variables that obey a particular distribution,
e.g.\ the uniform distribution $\pi_{\rm uni}$ given by
$w_{i,i\pm1}=1$ for $0<w_{i,i\pm1}<1$ and $0$ otherwise, or the binary
distribution
\be
\pi_{\rm bin}(w_{i,i-1})=\frac{1}{2}\delta(w_{i,i-1}-\lambda) 
+ \frac{1}{2}\delta(w_{i,i-1}-\lambda^{-1})\;,\quad w_{i,i+1}=1
\quad\forall i
\label{distr}
\ee
with $\lambda$ some arbitrary parameter. For the disordered case
physical observables have to be averaged over this distribution $\pi$
which is denoted by square brackets $[\ldots]_{\rm av}$. Note we
consider the general case of asymmetric hopping rates $w_{i,i+1}\ne
w_{i+1,i}$ and we do not confine ourselves to the so called random
force model with correlated transition probabilities parameterized as
$w_{i,i \pm 1}=A \exp(\pm \phi_i)$ with random, uncorrelated
potentials $\phi_i$ on each site.

In order to define {\it single walker} persistence probabilities we
put an adsorbing boundary at site $i=0$, which means that we set
$w_{1,0}=0$ and also introduce a finite size length scale $L$ into the
system by putting another adsorbing boundary at $i=L+1$, i.e.\ setting
$w_{L+1, L}=0$.  Now we define the length scale dependent single
walker persistence $P_{\rm pr}(L,t)$ to be the probability that a
walker does not cross its starting point (i.e.\ does not get trapped
at site $i=0$) within the time interval $t$. The following
arguments\cite{diffusion} will then lead us to a scaling form for
$P_{\rm pr}(L,t)$: {\bf a)} the typical time the walker needs to reach
the site $L+1$ scales like $t_{\rm typ}\sim L^2$ in the case with
symmetric transition rates and like $\ln t_{\rm typ}\sim \sqrt{L}$ in
the asymmetric case, the Sinai-model, {\bf b)} the asymptotic limit
$P_{\rm surv}(L)=\lim_{t\to\infty}P_{\rm pr}(t,L)=
\lim_{t\to\infty}p_{L+1}(t)$, which is what we call a survival
probability, behaves like $P_{\rm surv}(L)\sim L^{-1}$ in the
symmetric case and like $P_{\rm surv}(L)\sim L^{-\theta_a}$ with
$\theta=1/2$ in the Sinai-model. Hence we expect
\be
[P_{\rm pr}(L,t)]_{\rm av}\sim\left\{
\begin{array}{lcl}
L^{-\theta_s}\cdot\tilde{p}_s(L^2/t) & \quad & ({\rm symmetric\;case})\\
L^{-\theta_a}\cdot\tilde{p}_a(\sqrt{L}/ \ln t) 
& \quad & ({\rm asymmetric\;case})
\end{array}
\right.
\label{scaling}
\ee
where the persistence exponents are $\theta_s=1$ for the symmetric
case and $\theta_a=1/2$ for the asymmetric case, and the scaling
functions behave like $\tilde{p}_{s}(x)\to x^{\theta_{s}/2}$ and
$\tilde{p}_{a}(x)\to x^{2\theta_{a}}$ for $x\to \infty$ and
$\tilde{p}_{s/a}(x)\to const.$ for $x\to 0$. In the infinite system
size limit one thus has for persistence probability $P_{\rm pr}(t)$ in
the asymmetric case $P_{\rm pr}(t)=\lim_{L\to\infty} P_{\rm
  pr}(L,t)\propto \ln(t)^{-2{\theta}_{a}}$, a logarithmically slow
decay reminiscent of the critical dynamics of the surface
magnetization of the random transverse Ising chain\cite{dynamics}
(RTIC). This is not incidental: the equivalence of the surface
magnetization in the latter quantum spin chain and the survival
probabilities of random walks has been formulated the first time in
\cite{bigpaper} and further analogies between anomalous diffusion and
the RTIC have been uncovered recently \cite{diffusion,flm}.

\deffig{fg:persistence}{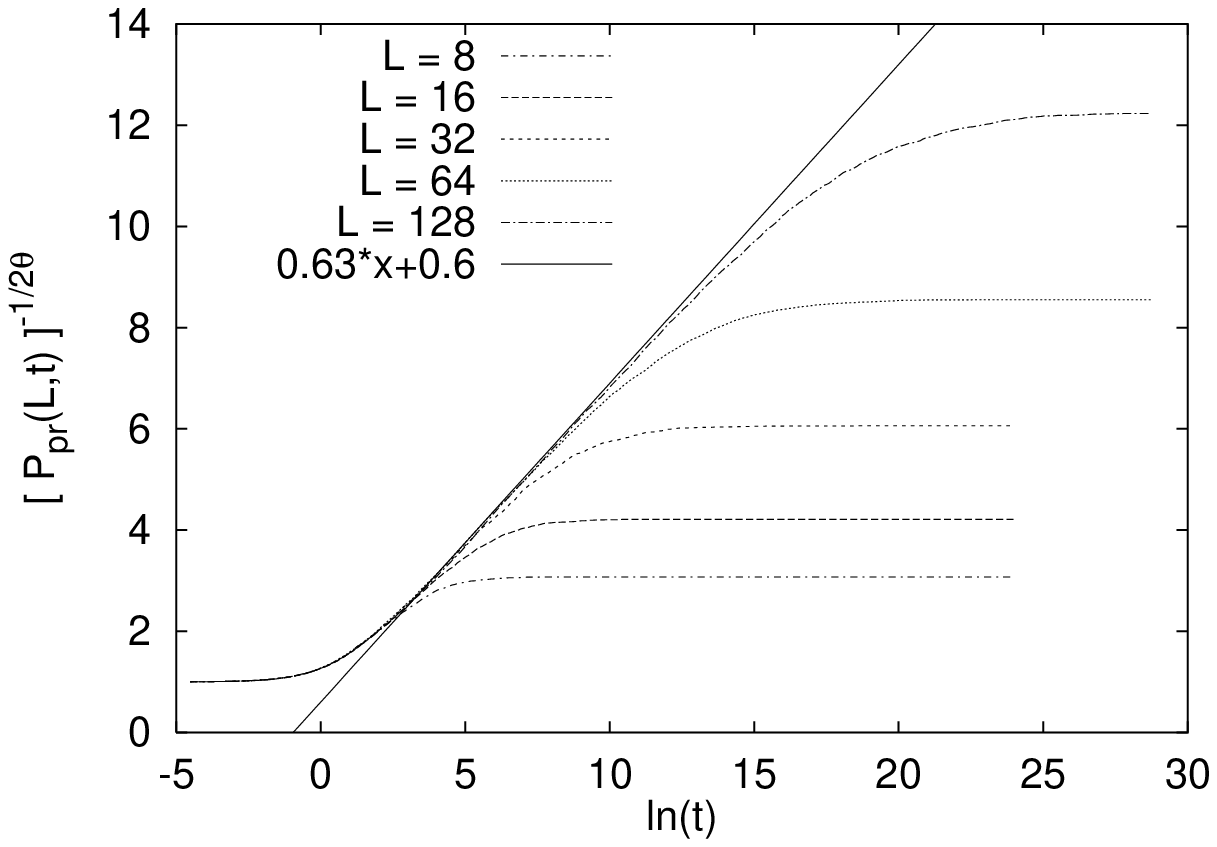}{63mm}{84mm} {The average persistence
  $[\overline{P}_{\rm pr}(L,t)]_{\rm av}$ to the power
  $-1/2\overline{\theta}_{a}$
versus $\ln t$, which should yield a straight line in the limit
$L\to\infty$, for different finite size scaling length $L$. The data
are for uniform distribution and averaged over 50.000 samples. We used
an estimate $\overline{\theta}_{a}=0.19$ to obtain an optimal data
collapse in the scaling region. The straight line indicates the
asymptotic result for the infinite system $L\to\infty$.}

It is known that diffusion in the Sinai model is different from normal
diffusion in many respects\cite{review}. Consider for instance one
particular disorder realization. Then an initially narrow probability
distribution of a walker peaked around, say, $x(t=0)=0$ does not
broaden with time, only its expectation value $\langle
x(t)\rangle=\sum_i i\cdot p_i(t)$ diffuses logarithmically slowly away
from its starting point. Therefore, in this situation in addition to
the single walker persistence one should also consider the persistence
properties of the average position of the walker $\langle
x(t)\rangle$, the {\it average} persistence $\overline{P}_{\rm
  pr}(t)$: This is the probability that up to a specified time $t$ the
average position of the walk in a particular environment has always
been on one side of the starting point, i.e.\ $\forall
0<t'<t:\;\langle x(t')\rangle>0$.

To study the average persistence we consider again a finite size
situation where we put an adsorbing boundary at $i=-L$ and one at
$i=L$ (note that now a single walker is not adsorbed when crossing the
starting point but only when he leaves the finite strip of width $2L$
centered around $i=0$). Then the average persistence
$\overline{P}_{\rm pr}(L,t)$ obeys the same scaling form as in the
second line of (\ref{scaling}), but the single walker persistence
exponent $\theta_{a}$ replaced by the average persistence exponent
$\overline{\theta}_{a}$:
\be
[\overline{P}_{\rm pr}(L,t)]_{\rm av}\sim
L^{-\overline{\theta}_a}\cdot\overline{p}_a(\sqrt{L}/ \ln t)\;.
\label{dissc}
\ee
Moreover, the persistence probability for the infinite system decays
again logarithmically slow: $\overline{P}_{\rm pr}(t)\propto
\ln(t)^{-2{\overline{\theta}}_{a}}$. Recently it has been conjectured
for the random force model\cite{flm} that the exponent
$\overline{\theta}_{a}$ is related to the golden mean via
$\overline{\theta}_{a}=(3-\sqrt{5})/4\approx0.191$.  In fig.\ 1 we
show numerical data that have been obtained by a numerical calculation
of $P_{\rm pr}(L,t)$ via diagonalization of the linear operator on the
r.h.s.\ of eq.(\ref{master}), which indicate that this might also hold
for the general asymmetric case.

In what follows we will demonstrate how to obtain a precise estimate
of the average persistence exponent $\overline{\theta}_{a}$.  Since in
the limit $t\to\infty$ the average persistence probability is given by
$\overline{P}_{\rm surv}\sim L^{-\overline{\theta}_{a}}$, the
computation of the survival probability $\overline{P}_{\rm surv}$ will
lead us to the desired result. We will now relate this survival
probability to a problem in the statistics of extreme events of
homogeneous random walks. To this end we consider the binary
distribution $\pi$ in (\ref{distr}) in the limit $\lambda\to 0$, which
means that one can discriminate between "forward" bonds $(i,i+1)$,
which are those for which $w_{i,i-1}=\lambda\to0$, i.e.\ those where
almost only (with probability $1/(1+\lambda)\to1$ jumps from $i$ to
$i+1$ occur, and "backward" bonds, which have
$w_{i,i-1}=\lambda^{-1}\to\infty$, implying almost always jumps from
$i$ to $i-1$ (again with probability
$\lambda^{-1}/(1+\lambda^{-1})\to1$).  This implies that when
sketching the transition rates as configuration as is done in fig.\ 2
the disorder configuration can be done visualized as a random
landscape with hills and valleys. Thus a walker starting at $i=0$ is
{\it on average} first driven to the first minimum $M_1$, and
considering the finite size situation where $L=L_1$ this disorder
configuration counts as a surviving configuration, since the walker in
average spends most of its time at $M_1$. Increasing the length scale
$L$ further beyond $L_2$ will then drive the average position of the
walker over the barrier $B$ into the valley $M_3$ at {\it negative}
coordinates, which means that this configuration is dead, i.e\ not
surviving.
  
\deffig{fg:sketch}{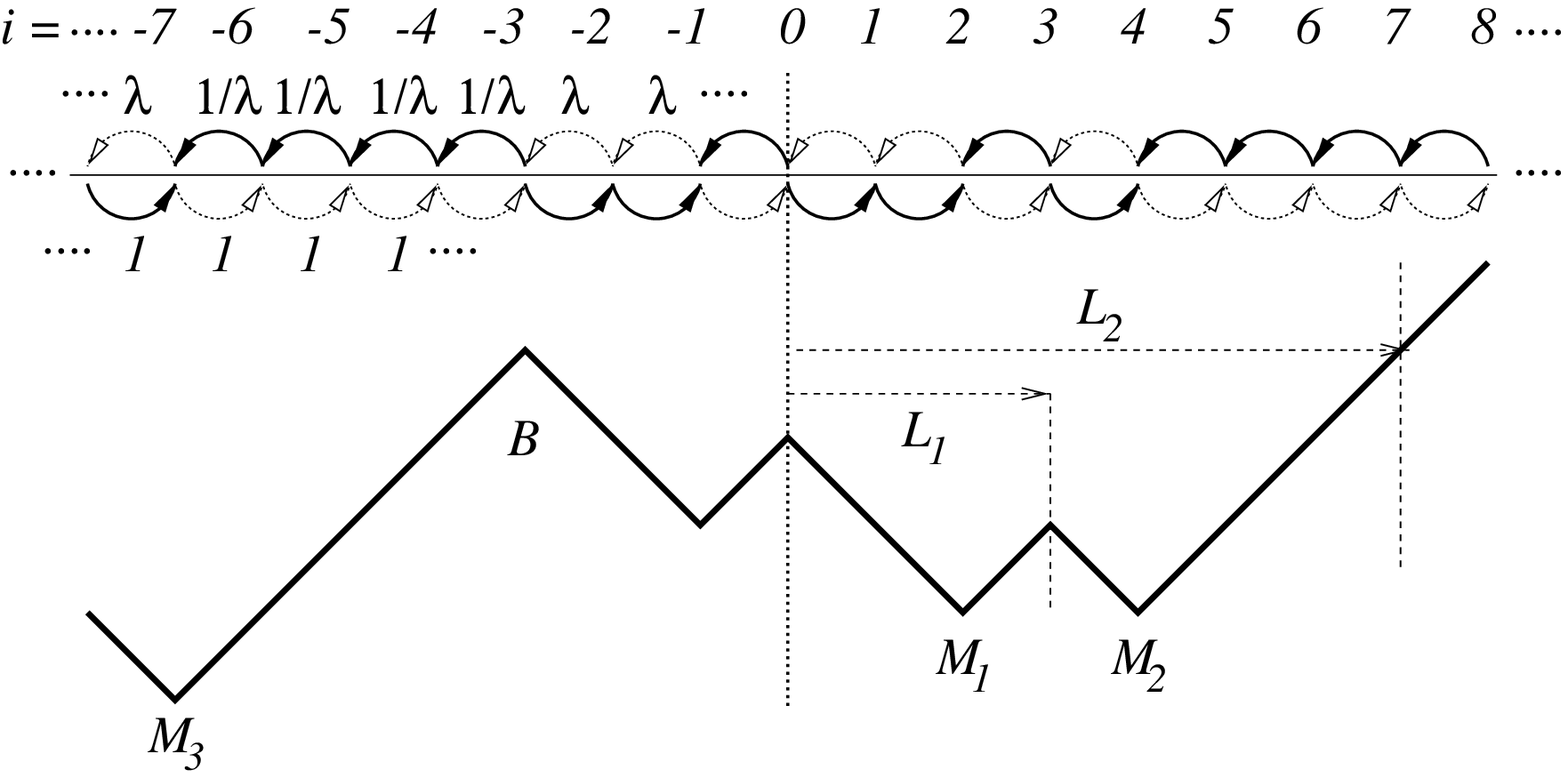}{50mm}{84mm}
{Sketch of the random landscape that corresponds to the binary
distribution (\protect{\ref{distr}}). In the limit $\lambda\to0$
transitions corresponding to broken arrows will rarely occur whereas
transitions corresponding to full arrows will occur with probability
close to one. For further details describing the effective dynamics
see text.}

Thus we conclude that the computation of $\overline{P}_{\rm surv}$
amounts to counting all surviving configurations of a random
landscape, i.e.\ a homogeneous random walk, in the manner described
above and the ratio of surviving configurations is just
$\overline{P}_{\rm surv}$. In other words we consider random walks of
length $2L$, corresponding to realizations of the transition
probabilities according to (\ref{distr}) in a strip of width $2L$
centered around the starting point, and say that it is in average
surviving if certain conditions are fulfilled. These conditions are
checked by inspecting the random landscape generated by the disorder
configuration (i.e\ the transition rates): one scans the landscape in
both directions from the starting point and denotes with $h(i)$
($i=-L,-L+1,\ldots,-1,0,1,2,\ldots,L$) the position of the walker (or
height of the landscape) at step (or site) $i$. We define the extreme
events to the right and to the left of $i=0$ by $x_{\rm
  max/min}(i)={\rm max/min}\{h(j)\vert 0\le j\le i\}$ and $y_{\rm
  max/min}(i)={\rm max/min}\{h(j)\vert -i\le j\le 0\}$, respectively,
and check iteratively (from $i=1$ to $i=L$) whether
\be
x_{\rm min}(i)> y_{\rm min}(i)\quad{\bf and}\quad
x_{\rm max}(i)> y_{\rm max}(i)\;.
\label{criterion}
\ee
If this happens for some site $i$, as it does for $i=8$ in fig.\ 2, it
means that there is a lower minimum on the left side of the starting
point ($x_{\rm min}<y_{\rm min}$) and that the walker can go there
since the barrier in between is low enough ($x_{\rm max}<y_{\rm max}$).
This implies that this configuration is dead.

In the inset of fig.\ 3 we show the results for a numerical estimate
of the survival probability $\overline{P}_{\rm surv}(L)$ by inspecting
10$^5$ random walk configuration for different system sizes --- the
data fit well to a power law with the exponent
$\overline{\theta}_{a}=0.19$. Next we implemented a recursive routine
that computes the number of surviving configuration according to the
above extreme events criterion (\ref{criterion}) {\it exactly}. Here
we took special care to the degenerate minima (like $M_1$ and $M_2$ in
fig.\ 2), in which case we assumed the arithmetic mean of the location
of the two to be the effective position ($L_1$ in fig.\ 2). From these
exact data for system sizes up to $L=14$ we can extract via
$\overline{\theta}_{a}(L_a,L_b)=-\ln(P_{\rm surv}(L_a)/P_{\rm
  surv}(L_b))/\ln(L_a/L_b)$ an effective finite size exponent that
approaches the exact value. In fig.\ 3 we show these exact data for
$L_a=L=L_b+1$ and $L_a=L=L_b+2$. Finally, from a standard series
extrapolation procedure\cite{BST} applied to these data we obtain our
best estimate which is
\be
\overline{\theta}_{a}=0.191\pm0.002\;,
\label{thetadis}
\ee
in very good agreement with the conjectured value 
$\overline{\theta}_{a}=(3-\sqrt{5})/4=0.19098\ldots$
for the random force model\cite{flm}.

\deffig{fg:exponent}{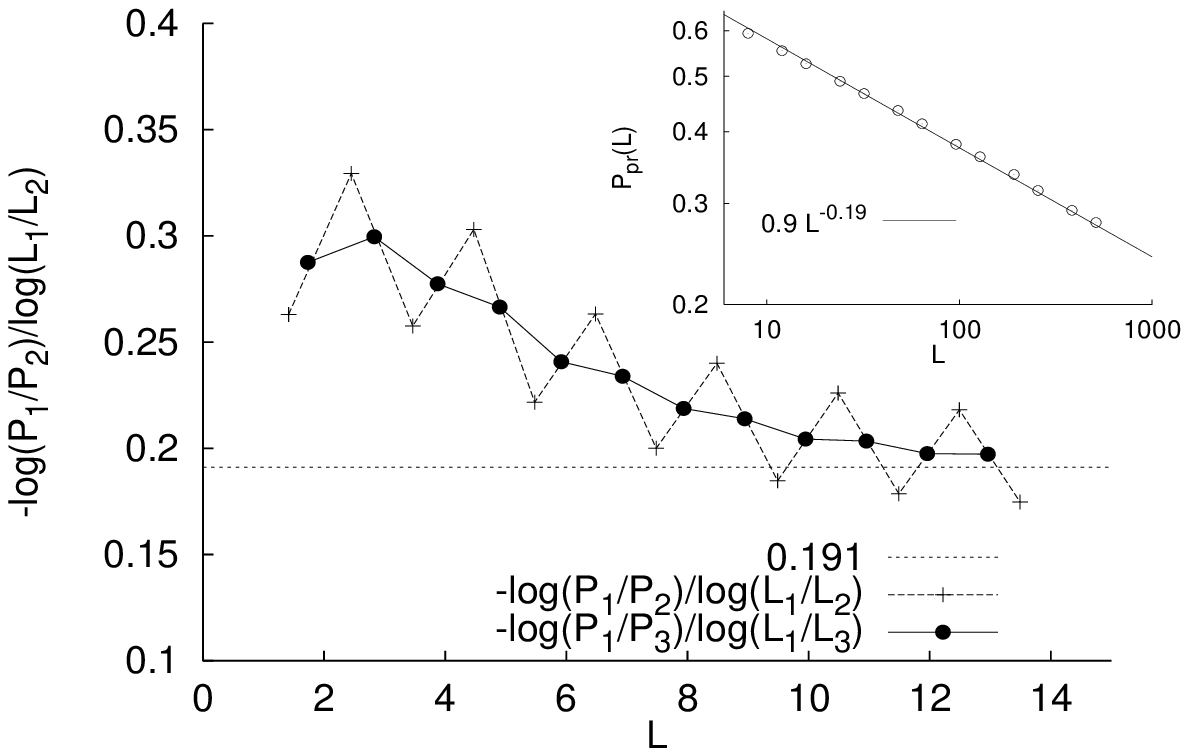}{63mm}{84mm} {The effective finite size
  exponent $\overline{\theta}_{a}$ from exact values for
  $\overline{P}_{\rm pr}(L)$ via complete enumeration of all surviving
  configuration. The asymptotic value $\overline{\theta}_{a}=0.191$
  obtained by series extrapolation methods is also shown. The insert
  shows the result for $\overline{P}_{\rm pr}(L)$ via stochastic
  enumeration for larger system sizes, giving
  $\overline{\theta}_{a}=0.19\pm0.01$.}

Although the concept of an average persistence seems to be based upon
the special feature of non-dispersive diffusion in the Sinai model,
which we described above, it turns out that there is an analogy to it
in normal diffusion, too. To show this, we first introduce the
integrated position of the walker at step $t$ as $I_t=\sum_{\tau=1}^t
i_{\tau}$, where $i_{\tau}$ is its position at time-step $\tau$. Then,
we define the {\it center of mass persistence} through the survival
condition for the integrated position as $I_{t'} \ge 0$ for $0<t'<t$.
For the disordered case we define a center of mass position via
\be
C(i)=\sum_{j=-i}^i j\cdot M(h(j))\;.
\label{CI}
\ee
where the weights $M(h(j))$, which are proportional to the average
time the walker spent at a given site, depend on the random
configurations.  In a strip $(-L \le i \le L)$ with adsorbing
boundaries the survival condition of the average position in the large
$t$-limit is $C(i)\ge0$ for $i=1,\ldots,L$, which is equivalent to the
survival condition in (\ref{criterion}).  For the extreme binary
distribution, we consider here, the weight function is singular, such
that $M(h')/M(h)\to0$ for $h'<h$, which is a consequence of the limit
$\lambda\to0$ for the distribution (\ref{distr}).  Since the form of
the distribution of the random transition rates is generally
irrelevant one expects that the scaling properties of the average
persistence and those of the center of mass persistence in (\ref{CI})
are equivalent for other type of distributions as well.

For a homogeneous walk one can also study the center of mass
persistence or the survival probability of the $I_t$ integrated
position. This type of problem has already been considered by
Sinai\cite{sinai2} for a discrete model and later by
Burkhardt\cite{burkhardt} for the continuum model. According to these
exact results in a homogeneous infinite system the long-time behavior
of the average persistence is given by $\lim_{L \to \infty}
\overline{P}_{pr}^{hom}(L,t) \sim t^{-1/4}$. We use this result to
write for finite systems the scaling conjecture:
\be
\overline{P}_{pr}^{hom}(L,t)=L^{-\overline{\theta}_{hom}}
\overline{p}_{hom}(L/t)\;
\label{homsc}
\ee
with $\overline{\theta}_{hom}=1/4$ and the scaling function behaves
like $\overline{p}_{hom}(x) \sim x^{\overline{\theta}_{hom}}$ for $x
\to \infty$ and $\overline{p}_{hom}(x) \to const$ for $x \to 0$. The
scaling combination, $x=L/t$ in (\ref{homsc}), follows from the
scaling properties of the Brownian motion. In $t$ steps the walker
typically crosses its starting position $\sim t^{1/2}$ times and its
displacement is typically of $\sim t^{1/2}$, thus the integrated
displacement is $\sim t$, from which the scaling relation $L \sim t$
follows. It is interesting to note that the integrated Brownian curve
is an {\it isotropic critical object}, with a correlation length
critical exponent $\nu=1$, as for the two-dimensional Ising model.

By an exact enumeration up to $L=30$ we have checked the relation in
(\ref{homsc}) in the limit $t \to \infty$ and got an estimate
$\overline{p}_{hom}(x=0)=0.6183(3)$ for the scaling function. It is
amusing to note that the above value agrees, within the accuracy of
the estimate, with the inverse of the golden mean ratio
$1/\tau=2/(\sqrt{5}+1)=0.6180$.

Concluding, we introduced the concept of average persistence in random
walks and studied its scaling properties in homogeneous and in
disordered environments. The finite-size scaling behavior of
$\overline{P}_{pr}(L,t)$ as given in (\ref{dissc}) and (\ref{homsc})
involve new exponents: $\overline{\theta}$ in (\ref{thetadis}) and
$\overline{\theta}_{hom}=1/4$ for the disordered and homogeneous
problems, respectively. It is interesting to note on an analogy
between the finite-size scaling form of the average persistence of
random walks and that of the magnetization autocorrelation function
$G(L,t)=[\langle \sigma_{L/2}^x(0) \sigma_{L/2}^x(t)]_{av}$ of the
transverse-field Ising spin chain of length $L$. According to
exact\cite{mccoy} and conjectured plus numerical\cite{bulkmag} results
we have for the finite-size scaling behavior\cite{dynamics}
\be
G(L,t)\sim\left\{
\begin{array}{lcl}
L^{-2x_{hom}}\cdot\tilde{g}_{hom}(L/t) & \quad & ({\rm homogeneous\;case})\\
L^{-x_{rand}}\cdot\tilde{g}_{rand}(\sqrt{L}/ \ln t) 
& \quad & ({\rm random\;case})
\end{array}
\right.
\label{mscaling}
\ee
with the magnetization scaling dimensions $x_{hom}=1/8$ and $x_{rand}=
(3-\sqrt{5})/2\approx 0.191$\cite{bulkmag}, for the homogeneous and
random transverse-field Ising models, respectively. Since $2 x_{hom}=
\overline{\theta}_{hom}$ and $x_{rand}=\overline{\theta}_a$ we have a
complete correspondence between the scaling behavior of the average
persistence of random walks and the magnetization autocorrelation
function of the transverse-field Ising chain, both in the homogeneous
and in the disordered problems. These relations then complete the
previously observed correspondences\cite{diffusion} between single
walker persistence and the surface autocorrelation function of the
transverse-field Ising spin chain.

This work has been supported by the Hungarian National Research Fund
under grants No OTKA TO23642, OTKA TO25139 and OTKA TO15786 and by the
Ministery of Education under grant No FKFP 0765/1997. H.\ R.'s work
was supported by the Deutsche Forschungsgemeinschaft (DFG).
Useful discussions with T.W. Burkhardt are gratefully acknowledged.

\end{document}